# Fuzzy sets in nonparametric Bayes regression

**Jean-François Angers[1] and Mohan Delampady[2]**

*Université de Montréal and Indian Statistical Institute, Bangalore*

**Abstract:** A simple Bayesian approach to nonparametric regression is described using fuzzy sets and membership functions. Membership functions are interpreted as likelihood functions for the unknown regression function, so that with the help of a reference prior they can be transformed to prior density functions. The unknown regression function is decomposed into wavelets and a hierarchical Bayesian approach is employed for making inferences on the resulting wavelet coefficients.

## Contents

## 1. Introduction

Consider the model

$$(1.1) \qquad y_i = g(x_i) + \varepsilon_i, \quad i = 1, \ldots, n, \text{ and } x_i \in \mathcal{T},$$

where $\varepsilon = (\varepsilon_1, \varepsilon_2, \ldots, \varepsilon_n)' \sim N(0, \sigma^2 I)$, $\sigma^2$ is unknown and $g(\cdot)$ is a function defined on some index set $\mathcal{T} \subset \mathcal{R}^1$. Inferences about $g$ such as its estimation and estimation error as well as model checking are of interest.

Without parametric assumptions such as those leading to linear regression, this is a nonparametric regression problem. A Bayesian approach to (fully) nonparametric regression problems typically requires specifying prior distributions on function spaces which is rather difficult to handle. The extent of the complexity of this approach can be gauged from sources such as Ghosh and Ramamoorthi [9], Lenk [10], and so on. Furthermore, quantifying useful prior information such as "$g$ is close to (a specified function) $g_0$" is difficult probabilistically, whereas this seems quite straightforward if instead an appropriate metric on the concerned function space is used. This is where fuzzy sets or membership functions can be made use of. Before proceeding to this problem, let us recall the following details on fuzzy sets and membership functions.

---

[1]Département de mathématiques et de statistique, Université de Montréal, Montréal H3C 3J7, Canada, e-mail: angers@dms.umontreal.ca

[2]Statistics and Mathematics Unit, Indian Statistical Institute, Bangalore 560 059, India, e-mail: mohan@isibang.ac.in

**Definition 1.1.** A fuzzy subset $A$ of a space $\mathcal{G}$ (or just a fuzzy set $A$) is defined by a membership function $h_A : \mathcal{G} \longrightarrow [0, 1]$.

The membership function, $h_A(g)$, is supposed to express the degree of compatibility of $g$ with $A$. For example, if $\mathcal{G}$ is the real line and $A$ is the set of points "close to 0", then $h_A(0) = 1$ indicates that 0 is certainly included in $A$, but $h_A(.05) = .01$ says that .05 is not really "close" to 0 in this context. Similarly, if $\mathcal{G}$ is a set of functions and $A \subset \mathcal{G}$ is a set of functions "close" to a given function $g_0$, then $h_A(g_0) = 1$ indicates that $g_0$ is certainly included in $A$; however, if $h_A(g_1) = .01$ with $g_1(x) = 10g_0(x) + 100$ then $g_1$ is not really "close" to $g_0$ in this case.

Note that even when $\mathcal{G} = \Theta$ is the parameter space, a membership function $h_A(\theta)$ is not a probability density or mass function defined on $\Theta$, and hence cannot be used to obtain a prior distribution directly. Instead, as we have done in [4], we propose that a reasonable interpretation for a fuzzy subset $A$ of $\Theta$ is that it is a likelihood function for $\theta$ given $A$. (See also [12, 13]). This interpretation seems to be able to answer some of the questions regarding how appropriate the concept of *fuzziness* is in modeling our perception of imprecision. French [7] discusses a few of these questions. First of all, likelihood is an accepted means for modeling imprecision. Another important question is how to define $h_{A \cap B}$ from $h_A$ and $h_B$ for incorporating $h_A$ and $h_B$ in Bayesian inference. If $A$ and $B$ are independent, then interpreting $h_A$ and $h_B$ as likelihood functions leads to the result that $h_{A \cap B} = h_A h_B$, for this purpose. Further, the qualitative ordering that underlies a membership function can also be investigated with this interpretation, in conjunction with a prior distribution, as we study later. See [4] for an application to hierarchical and robust Bayes inference.

This paper is closely related to Angers and Delampady [3, 4]. In the latter, we used fuzzy sets to help specify a prior density on a finite parameter space, whereas in the former a nonparametric function estimator using wavelet decomposition was presented. It is therefore natural to explore whether a combination of these two ideas can be fruitfully employed. Towards this end, we adopt the semi-parametric approach of using the wavelet decomposition of a regression function along with a remainder, thus effecting a drastic reduction of the dimension of the parameter space. Next we propose to incorporate the imprecise prior information of the kind "the true regression function $g$ is close to $g_0$" using membership functions. Wavelet decomposition of $g_0$ provides the wavelet coefficients $\theta_0$ which we take as the prior mean of the wavelet coefficients $\theta$ corresponding to $g$. Precision of this prior mean estimate is unclear, so we treat the prior variance of these coefficients as a hyper-parameter with a second stage (reference) prior. This approach is also tied to the question of Bayesian robustness. If a membership function stating "we think $g$ is close to $g_0$" is the only prior input that we intend to incorporate, how should we then proceed with the Bayesian inference related to $g$? We claim that the only natural approach is to study the robustness of the inferences resulting from the class of prior densities compatible with this membership function (see Section 5.1). This approach is further explained in the following sections.

This paper is organized as follows. In Sections 2 and 3, a brief summary of our previous work [3, 4] is presented. In Section 4, details on the posterior calculations are given. The main focus of this paper, namely, model checking using Bayes factors and fuzzy models, is discussed in Section 5. Simulations and practical examples are presented in the last section to illustrate this theme.

## 2. Nonparametric regression and wavelets

To proceed further, we assume that the regression function $g$ and its prior guess $g_0$ are in $\mathcal{L}_2$ and impose a wavelet structure on both of them. To do this, we consider a compactly supported wavelet function $\psi \in \mathcal{C}^s$, the set of real-valued functions with continuous derivatives up to order $s$. Thus, $g(x)$ can be written as

$$g(x) = \sum_{|k| \leq K_0} \alpha_k \phi_k(x) + \sum_{j=0}^{\infty} \sum_{|k| \leq K_j} \beta_{j,k} \psi_{j,k}(x)$$

$$= g_J(x) + R_J(x),$$

where

$$
\begin{aligned}
g_J(x) &= \sum_{|k| \leq K_0} \alpha_k \phi_k(x) + \sum_{j=0}^{J} \sum_{|k| \leq K_j} \beta_{j,k} \psi_{j,k}(x), \\
R_J(x) &= \sum_{j=J+1}^{\infty} \sum_{|k| \leq K_j} \beta_{j,k} \psi_{j,k}(x), \\
\phi_k(x) &= \phi(x-k), \\
\psi_{j,k}(x) &= 2^{j/2} \psi(2^j x - k), \\
\alpha_k &= \int_{\mathcal{T}} \phi(x-k) g(x) dx \text{ and} \\
\beta_{j,k} &= \int_{\mathcal{T}} 2^{j/2} \psi(2^j x - k) g(x) dx.
\end{aligned}
$$

(2.1)

Here $K_j$ is such that $\phi_k(x)$ and $\psi_{j,k}(x)$ vanish on $\mathcal{T}$ whenever $|k| > K_j$, and $\phi$ is the scaling function ("father wavelet") corresponding to the "mother wavelet" $\psi$. Such $K_j$'s exist (and are finite) since the wavelet function that we have chosen has compact support. (See [3] or [8] for details.) Since the number of observations is finite, only a finite number of parameters can be estimated. Therefore, as suggested in [3], the resolution level $J$ should be chosen such that the total number of unknown parameters which need to be estimated is no larger than $n$, the number of observations. Since the total number of $\alpha$ and $\beta$ parameters is bounded by

$$l_X 2^{J+1} + J(l_\psi + 1) + (l_\phi + l_\psi + 2)$$

where $l_X, l_\psi$ and $l_\phi$ represent the length of the support of $\mathcal{T}$, $\psi(\cdot)$ and $\phi(\cdot)$ respectively, we require this upper bound to be less than $n$. The optimal resolution level $J$, however, should be chosen using a Bayesian model selection technique, for example [3]. Consequently, we propose that the function $g_J(\cdot)$ be estimated from the data and the function $R_J(\cdot)$ be considered as a "nuisance parameter" to be eliminated by integrating it out.

Further we assume that

$$g_0(x) = \sum_{|k| \leq K_0} \alpha_{0,k} \phi_k(x) + \sum_{j=0}^{\infty} \sum_{|k| \leq K_j} \beta_{0,j,k} \psi_{j,k}(x),$$

where

$$\alpha_{0,k} = \int_{\mathcal{T}} \phi(x-k) g_0(x) dx,$$

$$\beta_{0,j,k} = \int_{\mathcal{T}} 2^{j/2} \psi(2^j x - k) g_0(x) dx.$$

Since $g_0$ is given, the wavelet coefficients can be assumed known.

From [1], it is known that $\beta_{j,k} = \mathrm{O}(2^{-js})$ where $s$ denotes the degree of "smoothness" of $\psi$, that is $\psi(\cdot) \in \mathcal{C}^s$ $(s > 1/2)$. Further, some of the *a priori* information wavelet coefficients can be translated into

(2.2) $\qquad\qquad E[\alpha_k] = \alpha_{0,k}; \qquad\qquad E[\beta_{j,k}] = \beta_{0,j,k};$

(2.3) $\qquad\qquad \mathrm{Var}(\alpha_k) = \tau^2; \qquad\qquad \mathrm{Var}(\beta_{j,k}) = \tau^2/2^{2js};$

(see [3] for details). Here, $\tau^2$ is a first stage hyper-parameter, a suitable prior on which will be specified later. (It is also supposed that $E[\beta_{j,k}] = \beta_{0,j,k} = 0$ for $j \geq J+1$.) However, the prior probability distribution on the coefficients $\alpha_k$ and $\beta_{j,k}$ itself will be specified by a membership function to be introduced later.

Since there are only a finite number of $\beta_{j,k}$ that can be estimated, we cannot expect to have a very informative prior distribution on the remainder term, $R_J$. However, to be able to proceed with the Bayesian approach, we need to assume that $R_J(\cdot)$ is a stochastic process. For simplicity and computational ease, we consider a Gaussian process with zero mean (compatible with $E[\beta_{j,k}] = \beta_{0,j,k} = 0$ for $j \geq J+1$). Specifically, following [3], we now assume that $R_J(x)$ is a Gaussian process with mean function 0 and covariance kernel $\tau^2 Q(x, y)$ where

(2.4) $\qquad\qquad Q(x, y) = \sum_{j=J+1}^{\infty} \frac{1}{2^{2js}} \sum_{|k| \leq K_j} \psi_{j,k}(x) \psi_{j,k}(y).$

Note that a prior assumption such as $\beta_{j,k}$ being independent normal random quantities will naturally lead to this prior distribution as can be seen from the Karhunen–Loeve expansion [2]. We, however, make the stronger assumption that the remainder $R_J(\cdot)$ itself is a zero-mean Gaussian but with an unknown prior variance component $\tau^2$. Our reason for this assumption is that, once the optimal resolution level $J$ is chosen using a powerful mechanism such as a Bayes factor, as we suggest later, the wavelet coefficients at higher resolutions are not expected to have substantial influence on the wavelet smoother. Hence $R_J(\cdot)$ which is made up of these higher-order wavelet coefficients will also have negligible influence.

## 3. Prior information and membership functions

We have explained in the previous section that we would like to make use of imprecise prior information such as "$g$ is close to $g_0$" by using a membership function which translates this into a measure of distance between the corresponding wavelet coefficients. Let us examine the implications of assuming that the available prior information is quantified in terms of a membership function

$$h_A(g) = \xi(\rho(g, g_0)), \tag{3.1}$$

where $\rho$ is a measure of distance. Due to the wavelet decomposition assumed on $g$ as well as $g_0$ (see Section 2), a natural choice for $\rho$ is the $\mathcal{L}_2$ distance given by

$$\rho^2(g, g_0) = ||g - g_0||^2 = \sum (\theta_{j,k} - \theta_{j,k}^0)^2, \tag{3.2}$$

where $\theta_{j,k}$ and $\theta_{j,k}^0$ are, respectively, the wavelet coefficients of $g$ and $g_0$. Using Parseval's identity, note that

$$\rho^2(g, g_0) = ||g - g_0||^2 = \sum_{|k| \le K_0} (\alpha_k - \alpha_{0,k})^2 + \sum_{j=0}^{\infty} \sum_{|k| \le K_j} (\beta_{j,k} - \beta_{0,j,k})^2.$$

Since we cannot estimate $\beta_{j,k}$ for $j = J + 1, \ldots,$ and because $\beta_{j,k} = \mathrm{O}(2^{-js})$ (see [1]) we then have

$$\rho^2(g, g_0) = \sum_{|k| \le K_0} (\alpha_k - \alpha_{0,k})^2 + \sum_{j=0}^{J} \sum_{|k| \le K_j} (\beta_{j,k} - \beta_{0,j,k})^2 \tag{3.3}$$

$$+ \sum_{j=J+1}^{\infty} \sum_{|k| \le K_j} (\beta_{j,k} - \beta_{0,j,k})^2$$

$$= \sum_{|k| \le K_0} (\alpha_k - \alpha_{0,k})^2 + \sum_{j=0}^{J} \sum_{|k| \le K_j} (\beta_{j,k} - \beta_{0,j,k})^2 + \sum_{j=J+1}^{\infty} \sum_{|k| \le K_j} \mathrm{O}(2^{-2js})$$

$$= \sum_{|k| \le K_0} (\alpha_k - \alpha_{0,k})^2 + \sum_{j=0}^{J} \sum_{|k| \le K_j} (\beta_{j,k} - \beta_{0,j,k})^2 + \mathrm{O}(2^{-2(J+1)s})$$

$$= \rho_J^2(g, g_0) + O(2^{-2(J+1)s}). \tag{3.4}$$

For this reason, to quantify the imprecise prior information, we will use a membership function that will depend only on $\rho_J^2(g, g_0)$. Some possibilities for $h_A$ are the following:

(i) The Gaussian membership function given by

$$h_A(g) = \exp(-\rho_J^2(g, g_0)) = \exp(-\alpha ||\theta - \theta^0||^2). \tag{3.5}$$

This membership function can be explained as follows. Suppose we have available some past data of the form

$$y_i^* = g(x_i^*) + \varepsilon_i, \quad i = 1, \ldots, n^*,$$

with $\varepsilon_i$ denoting i.i.d. normal errors, and suppose $g$ is estimated from this data by $\hat{g}$. Then the information in this data may be quantified using a membership function of the type

$$h_A(g) = \exp(-\alpha ||g - \hat{g}||^2) = \exp(-c \sum (\theta_j - \hat{\theta}_j)^2).$$

$g_0$ may then be identified with $\hat{g}$. If we have multiple past data sets, we may then have available $h_{A_1}(g) = \exp(-\alpha_1||g - \hat{g}_1||^2)$, $h_{A_2}(g) = \exp(-\alpha_2||g - \hat{g}_2||^2)$, and so on, which may be combined into

$$h_A(g) = h_{A_1 \cap A_2}(g) = h_{A_1}(g)h_{A_2}(g) = \exp(-\{\alpha_1||g - \hat{g}_1||^2 + \alpha_2||g - \hat{g}_2||^2\}).$$

As an example one could consider fitting regression lines to two (or more) sets of past data with possibly different error variances and use the fitted regression lines along with the estimated variances for constructing the membership functions. This justifies to some extent our previous suggestion that membership functions quantify prior information in the sense of the likelihood. The constants $\alpha_1$ and $\alpha_2$ provide additional scope for assigning different weights to the two sources of information, which is another appealing feature of this approach.

(ii) The multivariate $t$ membership function

$$(3.6) \quad h_A(g) = \left(1 + \rho_J^2(g, g_0)\right)^{-(p+q)/2} = \left(1 + (\theta - \theta^0)'V^{-1}(\theta - \theta^0)/q\right)^{-(p+q)/2},$$

where $q > 2$ is the degrees of freedom and $p$ denotes the dimension of $\theta$. This is a continuous scale mixture of Gaussian membership functions with the same $g_0$ for each of the membership functions. Since this vanishes more slowly than (3.5), one could expect better robustness with this.

(iii) The uniform function

$$h_A(g) = \begin{cases} 1, & \text{if } \rho_J(g, g_0) \le \delta; \\ 0, & \text{otherwise.} \end{cases}$$

This is an extreme case where $g$ is restricted to a neighborhood of $g_0$.

In order to proceed with Bayesian inference on $g$, we need to convert the membership function into a prior density. This is done as in [4] with the aid of a reference prior density $\pi_0$. Thus we obtain the prior density

$$\pi(g) \quad \propto \quad h_A(g)\pi_0(g),$$

or, upon utilizing the wavelet decomposition for $g$, we have an equivalent prior density

$$\pi(\theta, \sigma^2) \quad \propto \quad h_A(\theta)\pi_0(\theta, \sigma^2).$$

## 4. Posterior calculations

As in [3], let $Q_n = (Q_n)_{il}$, $1 \le i, l \le n$, where $(Q_n)_{il} = Q(x_i, x_l)$, which was introduced in (2.4). Note that

$$(Q_n)_{il} = \sum_{j \ge J+1} \sum_{|k| \le K_j} 2^{-2js} \psi_{jk}(x_i)\psi_{jk}(x_l).$$

Let $X = (\Phi', S')$ with the $i$th row of $\Phi'$ being $\{\phi_k(x_i)\}'_{|k| \le K_0}$ and the $i$th row of $S'$ being $\{\psi_{jk}(x_i)\}'_{|k| \le K_j, 0 \le j \le J}$. Then we have the model

$$(4.1) \qquad\qquad \mathbf{y}|\theta, \sigma^2, \tau^2 \sim N(X\theta, \sigma^2 I_n + \tau^2 Q_n).$$

Unless $\theta$ has a normal prior distribution or a hierarchical prior with a conditionally normal prior distribution, analytical simplifications in the computation of posterior quantities are not expected. For such cases, we have the joint posterior density of the wavelet coefficients $\theta$ and the error variances $\sigma^2$ and $\tau^2$ given by the expression

$$\pi(\theta, \sigma^2, \tau^2 | \mathbf{y}) \quad \propto \quad f(\mathbf{y} | \theta, \sigma^2, \tau^2) h_A(\theta) \pi_0(\theta, \sigma^2, \tau^2),$$

where $f$ is the likelihood. From (4.1), $f$ can be expressed as

$$
\begin{aligned}
f(\mathbf{y} | \theta, \sigma^2, \tau^2) \quad \propto \quad & |\sigma^2 I_n + \tau^2 Q_n|^{-1/2} \\
& \times \exp(-\tfrac{1}{2}\{(\mathbf{y} - X\theta)'(\sigma^2 I_n + \tau^2 Q_n)^{-1}(\mathbf{y} - X\theta)\}).
\end{aligned}
$$

Proceeding further, suppose $\pi_0$ of the form

$$(4.2) \qquad \pi_0(\theta, \sigma^2, \tau^2) = \pi_1(\sigma^2, \tau^2),$$

which is constant in $\theta$, is chosen.

MCMC based approaches to posterior computations are now readily available. For example, Gibbs sampling is straightforward. Note that the conditional posterior densities are given by

$$(4.3) \qquad \pi(\theta | \mathbf{y}, \sigma^2, \tau^2) \propto \exp(-\tfrac{1}{2}\{(\mathbf{y} - X\theta)'(\sigma^2 I_n + \tau^2 Q_n)^{-1}(\mathbf{y} - X\theta)\}) h_A(\theta),$$

$$(4.4)
\begin{aligned}
\pi(\sigma^2 | \mathbf{y}, \theta, \tau^2) \propto & |\sigma^2 I_n + \tau^2 Q_n|^{-1/2} \\
& \exp(-\tfrac{1}{2}\{(\mathbf{y} - X\theta)'(\sigma^2 I_n + \tau^2 Q_n)^{-1}(\mathbf{y} - X\theta)\}) \pi_1(\sigma^2, \tau^2),
\end{aligned}
$$

$$(4.5)
\begin{aligned}
\pi(\tau^2 | \mathbf{y}, \theta, \sigma^2) \propto & |\sigma^2 I_n + \tau^2 Q_n|^{-1/2} \\
& \exp(-\tfrac{1}{2}\{(\mathbf{y} - X\theta)'(\sigma^2 I_n + \tau^2 Q_n)^{-1}(\mathbf{y} - X\theta)\}) \pi_1(\sigma^2, \tau^2).
\end{aligned}
$$

However, major simplifications are possible with the Gaussian $h_A$ as in (i). In this case, the posterior analysis can proceed along the lines of [3]. Specifically, assuming that $h_A(\theta)$ is proportional to the density of $N(\theta_0, \tau^2 \Gamma)$ with

$$\Gamma = \begin{pmatrix} I_{2K_0+1} & 0 \\ 0 & \Delta_{M_\beta} \end{pmatrix},$$

where $M_\beta = \sum_{j=0}^{J}(2K_j+1)$ and with $\tau^2 \Delta$ being the variance-covariance matrix of $\beta$ (which is also diagonal, having the diagonal entries specified by $\mathrm{Var}(\beta_{jk}) = \tau^2/2^{2js}$), we obtain

$$(4.6)
\begin{aligned}
\mathbf{y} | \theta, \sigma^2, \tau^2 \quad &\sim \quad N(X\theta, \sigma^2 I_n + \tau^2 Q_n), \\
\theta | \tau^2 \quad &\sim \quad N(\theta_0, \tau^2 \Gamma).
\end{aligned}
$$

Therefore, it follows that

$$(4.7) \qquad \mathbf{y} | \sigma^2, \tau^2 \quad \sim \quad N(X\theta_0, \sigma^2 I_n + \tau^2 (X\Gamma X' + Q_n)),$$

$$(4.8) \qquad \theta | \mathbf{y}, \sigma^2, \tau^2 \quad \sim \quad N(\theta_0 + A(\mathbf{y} - X\theta_0), B),$$

where

$$
\begin{aligned}
A \quad &= \quad \tau^2 \Gamma X' \left( \sigma^2 I_n + \tau^2 (X\Gamma X' + Q_n) \right)^{-1}, \\
B \quad &= \quad \tau^2 \Gamma - \tau^4 \Gamma X' \left( \sigma^2 I_n + \tau^2 (X\Gamma X' + Q_n) \right)^{-1} X\Gamma.
\end{aligned}
$$

Now proceeding as in [3], we employ spectral decomposition to obtain $X\Gamma X' + Q_n = HDH'$, where $D = \text{diag}(d_1, d_2, \ldots, d_n)$ is the matrix of eigenvalues and $H$ is the orthogonal matrix of eigenvectors. Thus,

$$\sigma^2 I_n + \tau^2 \left( X\Gamma X' + Q_n \right) = H \left( \sigma^2 I_n + \tau^2 D \right) H' = \sigma^2 H \left( I_n + uD \right) H',$$

where $u = \tau^2/\sigma^2$. Then, the first stage (conditional) marginal density of $\mathbf{y}$ given $\sigma^2$ and $u$ can be written as

$$
\begin{aligned}
m(\mathbf{y} \mid \sigma^2, u) &= \frac{1}{(2\pi\sigma^2)^{n/2}} \frac{1}{\det(I_n + uD)^{1/2}} \\
&\quad \times \exp\left\{ -\frac{1}{2\sigma^2} (\mathbf{y} - X\theta_0)' H(I_n + uD)^{-1} H'(\mathbf{y} - X\theta_0) \right\} \\
&= \frac{1}{(2\pi\sigma^2)^{n/2}} \frac{1}{\prod_{i=1}^{n}(1 + ud_i)^{1/2}} \exp\left\{ -\frac{1}{2\sigma^2} \sum_{i=1}^{n} \frac{s_i^2}{1 + ud_i} \right\},
\end{aligned}
$$

(4.9)

where $\mathbf{s} = (s_1, \ldots, s_n)' = H'(\mathbf{y} - X\theta_0)$. We choose the prior on $\sigma^2$ and $u = \tau^2/\sigma^2$ qualitatively similar to that used in [3]. Specifically, we take $\pi_1(\sigma^2, u)$ to be proportional to the product of an inverse gamma density $\{k^{c-1}/\Gamma(c-1)\} \exp(-k/\sigma^2)(\sigma^2)^{-c}$ for $\sigma^2$ and the density of a $F(b, a)$ distribution for $u$ (for suitable choice of $k$, $c$, $a$ and $b$). Conditions apply on $a$ and $b$ as indicated in [3].

Once $\pi_1(\sigma^2, u)$ is chosen as above, we obtain the posterior mean and covariance matrix of $\theta$ as in the following result.

**Theorem 4.1.**

$$E(\theta|\mathbf{y}) = \theta_0 + \Gamma X'HE \left[ (I_n + uD)^{-1} \mid \mathbf{y} \right] s, \tag{4.10}$$

*where the expectation is taken with respect to*

$$\pi_{22}(u \mid \mathbf{y}) \propto \frac{u^{b/2}}{(a + bu)^{(a+b)/2}} \left( \prod_{i=1}^{n}(1 + ud_i) \right)^{-1/2} \left( 2k + \sum_{i=1}^{n} \frac{s_i^2}{1 + ud_i} \right)^{-(n+2c)/2}. \tag{4.11}$$

*and*

$$
\begin{aligned}
\text{Var}(\theta \mid \mathbf{y}) &= \frac{1}{n + 2c} E \left[ 2k + \sum_{i=1}^{n} \frac{s_i^2}{1 + ud_i} \mid \mathbf{y} \right] \Gamma \\
&\quad - \frac{1}{n + 2c} \Gamma X'HE \left[ \left( 2k + \sum_{i=1}^{n} \frac{s_i^2}{1 + ud_i} \right) (I_n + uD)^{-1} \mid \mathbf{y} \right] H'X\Gamma \\
&\quad + E \left[ M(u)M(u)' \mid \mathbf{y} \right],
\end{aligned}
\tag{4.12}
$$

*where $M(u) = \Gamma X'H(I_n + uD)^{-1}s$.*

The proof of Theorem 4.1 readily follows upon using standard hierarchical Bayesian model techniques (see [8], Section 9.1). The second part of Equation (4.10) can be viewed as a correction term added to the prior guess after observing the data $\mathbf{y}$.

Coming to the multivariate $t$ form of $h_A$ as in (ii), we note that the multivariate $t$ density is a continuous scale mixture of multivariate normal densities as, for example, shown in [11], i.e., $\theta$ has the multivariate $t$ distribution with location $\theta_0$ and scale matrix $V$ with density of the form (3.6) if and only if

$$\theta \mid \delta^2 \sim N(\theta_0, q\delta^2 V), \qquad (\delta^2)^{-1} \sim \chi_q^2.$$

This implies that with a $\pi_0$ as in (4.2), we obtain a hierarchical normal prior structure for $\theta$ with an additional hyper-parameter $\delta^2$. Consequently, we can proceed with the posterior computations exactly as with the Gaussian $h_A$, except that we will have a two-dimensional integration after we simplify our calculations as above. However, MCMC techniques can easily handle this case.

Computations with the $h_A$ given in (iii) are more difficult. Analytical simplifications as shown for the cases (i) and (ii) are not available here. Therefore, we utilize the MCMC computational scheme outlined in (4.3)–(4.5) above. Alternatively, the Metropolis–Hastings (M-H) algorithm may be employed.

## 5. Model checking and Bayes factors

An important and useful model checking problem in the present setup is checking the two models

$$M_0 : g = g_0 \qquad \text{versus} \qquad M_1 : g \neq g_0.$$

Under $M_1$, $(g = g(\theta), \tau^2, \sigma^2)$ is given the prior $h_A(\theta)\pi_0(\theta, \tau^2, \sigma^2)I(g \neq g_0)$, whereas under $M_0$, $\pi_0(\sigma^2)$ induced by $\pi_0(\theta, \tau^2, \sigma^2)$ is the only part needed. In order to conduct the model checking, we compute the Bayes factor, $B_{01}$, of $M_0$ relative to $M_1$:

$$B_{01}(\mathbf{y}) = \frac{m(\mathbf{y}|M_0)}{m(\mathbf{y}|M_1)}, \tag{5.1}$$

where $m(\mathbf{y}|M_i)$ is the predictive (marginal) density of $\mathbf{y}$ under model $M_i$, $i = 0, 1$. We have

$$m(\mathbf{y}|M_0) = \int f(\mathbf{y}|g_0, \sigma^2)\pi_0(\sigma^2)\, d\sigma^2$$

and

$$m(\mathbf{y}|M_1) = \int f(\mathbf{y}|\theta, \tau^2, \sigma^2)h_A(\theta)\pi_0(\theta, \tau^2, \sigma^2)\, d\theta\, d\tau^2\, d\sigma^2.$$

Since improper priors can lead to difficulties in model checking problems, here we must employ proper priors. (Note that for estimation purposes, the noninformative, improper prior $(\sigma^2)^{-c}$ corresponding to $k = 0$ would have worked in the previous section.) We will develop the methodology here for the Gaussian membership function only; the other cases are similar but computationally more intensive.

Recall from the previous section that in the case of a Gaussian membership function $h_A$, the posterior analysis is similar to that discussed in [3], and for the same reason, the computation of the Bayes factor is also similar to what was discussed there. As in the previous section $\pi_0(\theta, \tau^2, \sigma^2)$ will be constant in $\theta$, while $\sigma^2$ is inverse gamma and is independent of $v = \sigma^2/\tau^2$ which is given the $F_{a,b}$ prior distribution. (Equivalently, $u = 1/v = \tau^2/\sigma^2$ is given the $F_{b,a}$ prior as before.) Specifically, $\pi_0(\sigma^2) = \{k^{c-1}/\Gamma(c-1)\}\exp(-k/\sigma^2)(\sigma^2)^{-c}$, where $c$ and $k$ (small) are suitably chosen. Therefore,

$$m(\mathbf{y}|M_0) = \int f(\mathbf{y}|g_0, \sigma^2)\pi_0(\sigma^2)\, d\sigma^2 = (2\pi)^{-n/2}\frac{k^{c-1}}{\Gamma(c-1)}$$

$$\Gamma(n/2 + c - 1)\left\{k + \frac{1}{2}\sum_{i=1}^{n}(y_i - g_0(x_i))^2\right\}^{-(n/2+c-1)}.$$

Further, using (4.7), it follows that

$$(5.2) \quad m(\mathbf{y}|M_1, \sigma^2, u) = (2\pi\sigma^2)^{-n/2} \prod_{i=1}^{n} (1 + ud_i)^{-1/2} \exp\left\{-\frac{1}{2\sigma^2} \sum_{i=1}^{n} \frac{s_i^2}{1 + ud_i}\right\},$$

where $\mathbf{s} = (s_1, \ldots, s_n)' = H'(\mathbf{y} - X\theta_0)$ as before. Therefore,

$$
\begin{aligned}
m(\mathbf{y}|M_1) &= \int m(\mathbf{y}|M_1, \sigma^2, u)\pi_0(\sigma^2, u)\, d\sigma^2\, du \\
&= (2\pi)^{-n/2} \frac{k^{c-1}}{\Gamma(c-1)} \int \prod_{i=1}^{n} (1 + ud_i)^{-1/2} \pi_0(u) \\
&\qquad \left\{ \int \exp\left\{ -\frac{1}{\sigma^2}\left(k + \frac{1}{2}\sum_{i=1}^{n}\frac{s_i^2}{1 + ud_i}\right)\right\} (\sigma^2)^{-(n/2+c)}\, d\sigma^2 \right\}\, du
\end{aligned}
$$

$$
\begin{aligned}
(5.3) \qquad &= (2\pi)^{-n/2} \frac{k^{c-1}}{\Gamma(c-1)} \Gamma(n/2 + c - 1) \\
&\qquad \int \left\{ k + \frac{1}{2}\sum_{i=1}^{n}\frac{s_i^2}{1 + ud_i}\right\}^{-(n/2+c-1)} \prod_{i=1}^{n}(1 + ud_i)^{-1/2}\pi_0(u)\, du,
\end{aligned}
$$

where $\pi_0(u)$ denotes the $F_{b,a}$ density of $u$. Note that this involves only a straightforward single-dimensional integration. The resulting Bayes factor is illustrated later in our examples.

### 5.1. Prior robustness of Bayes factors

Note that the most informative part of the prior density that we have used is contained in the membership function $h_A$. Since a membership function $h_A(\theta)$ is to be treated only as a likelihood for $\theta$, any constant multiple $ch_A(\theta)$ also contributes the same prior information about $\theta$. Therefore, a study of the robustness of the Bayes factor that we obtained above with respect to a class of priors compatible with $h_A$ is of interest. Here we consider a sensitivity study using the *density ratio class* defined as follows. Since the prior $\pi$ that we use has the form $\pi(\theta, \tau^2, \sigma^2) \propto h_A(\theta)\pi_0(\theta, \tau^2, \sigma^2)$, we consider the class of priors

$$\mathcal{C}_A = \left\{\pi : c_1 h_A(\theta)\pi_0(\theta, \tau^2, \sigma^2) \leq \alpha\pi(\theta, \tau^2, \sigma^2) \leq c_2 h_A(\theta)\pi_0(\theta, \tau^2, \sigma^2), \alpha > 0\right\},$$

for specified $0 < c_1 < c_2$. We would like to investigate how the Bayes factor (5.1) behaves as the prior $\pi$ varies in $\mathcal{C}_A$. We note that for any $\pi \in \mathcal{C}_A$, the Bayes factor $B_{01}$ has the form

$$B_{01} = \frac{\int f(\mathbf{y}|g_0, \sigma^2)\pi(\theta, \tau^2, \sigma^2)\, d\theta\, d\tau^2\, d\sigma^2}{\int f(\mathbf{y}|\theta, \tau^2, \sigma^2)\pi(\theta, \tau^2, \sigma^2)\, d\theta\, d\tau^2\, d\sigma^2}.$$

Even though the integration in the numerator above need not involve $\theta$ and $\tau^2$, we do so to apply the following result (see [8], Theorem 3.9, or [4], Theorem 4.1).

Consider the density-ratio class

$$\Gamma_{DR} = \{\pi : L(\eta) \leq \alpha\pi(\eta) \leq U(\eta) \text{ for some } \alpha > 0\},$$

for specified non-negative functions $L$ and $U$. Further, let $q \equiv q^+ + q^-$ be the usual decomposition of $q$ into its positive and negative parts, i.e., $q^+(u) = \max\{q(u), 0\}$ and $q^-(u) = -\max\{-q(u), 0\}$. Then we have the following theorem (see [6]).

**Theorem 5.1.** *For functions $q_1$ and $q_2$ such that $\int |q_i(\eta)| U(\eta)\, d\eta < \infty$, for $i = 1, 2$, and with $q_2$ positive a.s. with respect to all $\pi \in \Gamma_{DR}$,*

$$\inf_{\pi \in \Gamma_{DR}} \frac{\int q_1(\eta) \pi(\eta)\, d\eta}{\int q_2(\eta) \pi(\eta)\, d\eta} \text{ is the unique solution } \lambda \text{ of}$$

$$(5.4) \qquad \int (q_1(\eta) - \lambda q_2(\eta))^- U(\eta)\, d\eta + \int (q_1(\eta) - \lambda q_2(\eta))^+ L(\eta)\, d\eta = 0,$$

$$\sup_{\pi \in \Gamma_{DR}} \frac{\int q_1(\eta) \pi(\eta)\, d\eta}{\int q_2(\eta) \pi(\eta)\, d\eta} \text{ is the unique solution } \lambda \text{ of}$$

$$(5.5) \qquad \int (q_1(\eta) - \lambda q_2(\eta))^+ U(\eta)\, d\eta + \int (q_1(\eta) - \lambda q_2(\eta))^- L(\eta)\, d\eta = 0.$$

We shall discuss this result for the Gaussian membership function only. Then, since the prior $\pi$ that we use has the form $\pi(\theta, \tau^2, \sigma^2) \propto h_A(\theta) \pi_0(\tau^2, \sigma^2)$, and we don't intend to vary $\pi_0(\tau^2, \sigma^2)$ in our analysis, we redefine $\mathcal{C}_A$ as

$$\mathcal{C}_A = \left\{ \pi(\theta) : c_1 h_A(\theta) \leq \alpha \pi(\theta) \leq c_2 h_A(\theta), \alpha > 0 \right\},$$

for specified $0 < c_1 < c_2$. Now, we re-express $B_{01}$ as

$$B_{01}(\pi) = \frac{\int \left\{ \int f(\mathbf{y}|g_0, \sigma^2) \pi_0(\sigma^2) d\sigma^2 \right\} \pi(\theta)\, d\theta}{\int \left\{ \int f(\mathbf{y}|\theta, \tau^2, \sigma^2) \pi_0(\tau^2, \sigma^2)\, d\tau^2\, d\sigma^2 \right\} \pi(\theta)\, d\theta} = \frac{\int q_1(\theta) \pi(\theta)\, d\theta}{\int q_2(\theta) \pi(\theta)\, d\theta},$$

where

$$q_1(\theta) \equiv \int f(\mathbf{y}|g_0, \sigma^2) \pi_0(\sigma^2) d\sigma^2 = m(\mathbf{y}|M_0),$$

$$q_2(\theta) = \int f(\mathbf{y}|\theta, \tau^2, \sigma^2) \pi_0(\tau^2, \sigma^2)\, d\tau^2\, d\sigma^2.$$

Then, Theorem 5.1 is readily applicable, and we obtain

**Theorem 5.2.**

$$\inf_{\pi \in \mathcal{C}_A} B_{01}(\pi) \text{ is the unique solution } \lambda \text{ of}$$

$$(5.6) \qquad c_2 \int (q_1(\theta) - \lambda q_2(\theta))^- h_A(\theta)\, d\theta + c_1 \int (q_1(\theta) - \lambda q_2(\theta))^+ h_A(\theta)\, d\theta = 0,$$

$$\sup_{\pi \in \mathcal{C}_A} B_{01}(\pi) \text{ is the unique solution } \lambda \text{ of}$$

$$(5.7) \qquad c_2 \int (q_1(\theta) - \lambda q_2(\theta))^+ h_A(\theta)\, d\theta + c_1 \int (q_1(\theta) - \lambda q_2(\theta))^- h_A(\theta)\, d\theta = 0.$$

## 6. Examples, simulations and illustrations

Illustrative examples, simulated as well as those involving real-life data, are discussed below. In all examples we have used two membership functions:

1. Gaussian membership function: $h_A(g)$ proportional to the density of $N(\theta_0, \tau^2\Gamma)$, where $\theta_0$ is obtained from the wavelet decomposition of $g_0$. The hyper-parameters $a$ and $b$ (see (4.11) and (4.12)) are $b = 3$ and $a = 8(b+2)/(b-2)$. Sensitivity analysis shows that the values of these hyper-parameters do not influence the results very much. To show that the other hyper-parameters $c$ and $k$ (see (4.11) and (4.12)) have some effect, though not substantial, we have displayed the wavelet smoothers (see Figures 1(a), 2(a) and 3(a)) for $(c, k) = (2, 1.5), (1.5, 0.5)$ and $(1.05, 0.05)$.
2. Uniform on the ellipsoid (see (iii) in Section 3): $h_A(g) = I_{\{\rho_J(g, g_0) \leq \delta\}}$. We have used three different values (0.5, 1 and 5) for $\delta$.

For the simulated examples, we generated observations from the model (1.1) with the regression function $g(x) = \cos(2\pi x)$ where $x$ is drawn from a uniform density on the unit interval. Then we considered three different prior guesses for $g_0$: (i) $g_0(x) = \cos(2\pi x)$ (see Figure 1), (ii) $g_0(x) = 4|x - 0.5| - 1$ (see Figure 2), and (iii) $g_0 \equiv 0$ (see Figure 3).

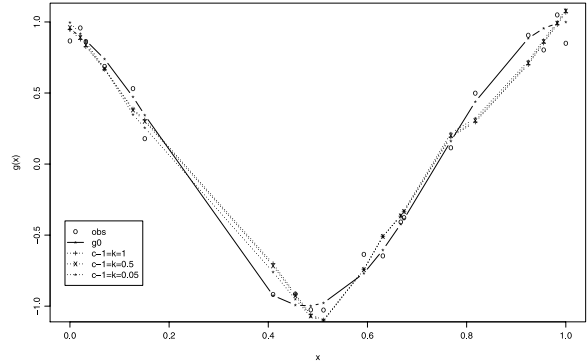

(a) Gaussian

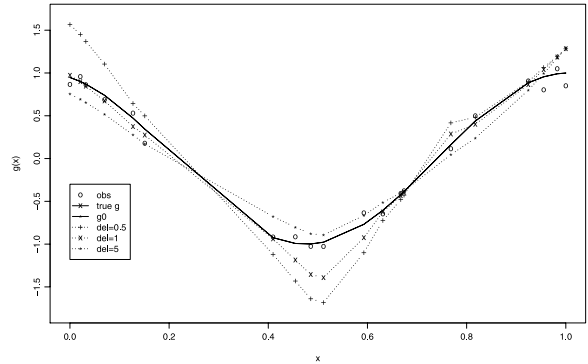

(b) Ellipsoid

Fig 1. *Wavelet smoother for $g(x) = \cos(2\pi x)$ and $g_0(x) = \cos(2\pi x)$.*

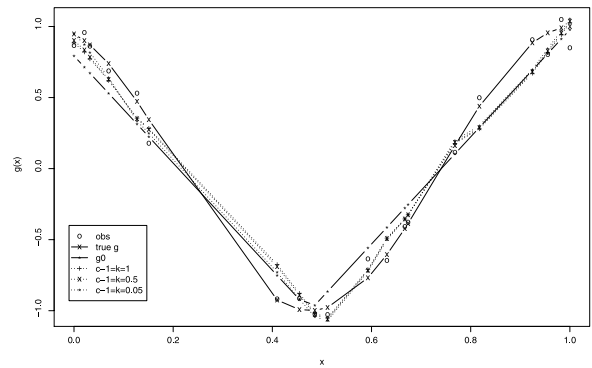

(a) Gaussian

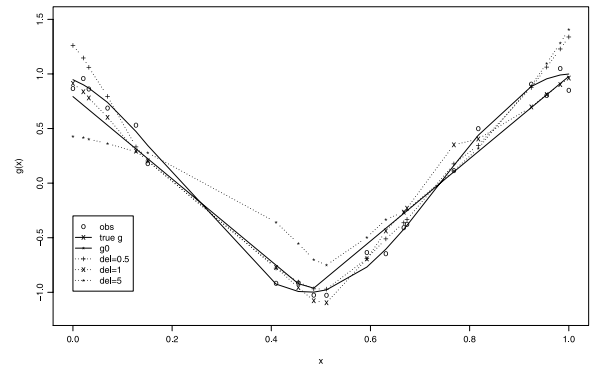

(b) Ellipsoid

Fig 2. *Wavelet smoother for* $g(x) = \cos(2\pi x)$ *and* $g_0(x) = 4|x - 0.5| - 1$.

Note that in (i) the chosen $g_0$ is the best possible prior guess. Further, since the normal prior is very informative, as expected the smoother (see Figure 1(a)) does an excellent job of extracting the true regression function. The behavior of the smoother obtained from the ellipsoid membership function (see Figure 1(b)) is similar to that seen in Figure 1 (a), even though the prior is different.

The smoother presented in Figure 2 behaves very similar to what was seen in Figure 1. The prior guess, $g_0$, is slightly different from the true $g$.

The behavior seen in Figure 3 emphasizes our comment following Figure 2. In fact, the smoother here looks better than the one in Figure 2. This is perhaps because the prior is less informative (concentrated) than the normal prior used there, so that the smoother can follow the data more closely than the prior.

We next applied our wavelet smoother to the "Humidity data" example from (see [3]; also [8], Example 10.2). The variable of interest $y$ that we have chosen from the data set is the weekly average humidity level. The observations were made from June 1, 1995 to December 13, 1998. We have chosen time (day of recording the observation) as the covariate $x$. Since we have 185 observations here, the maximum possible value for $J$ is 6.

In this data (see Figure 4(a)) a seasonality effect is present, so we have chosen $g_0(x) = 22.5 \cos(2\pi(x+0.1)/0.2) + 62.5$, where $x = $ (day − June 1, 1995)/(December 13, 1998 − June 1, 1995). This choice of $g_0$ is rather arbitrary but it seems to follow the seasonal variations. For the analysis which led to Figure 4(a) the Gaussian membership function was used while in Figure 4(b), it was the ellipsoid one. In this latter figure, we have also added the lower and upper envelopes obtained from the prior (labeled Min/Max). From these two figures, it can be seen that the proposed estimator fits the data well and the final result does not depend much on the membership function used.

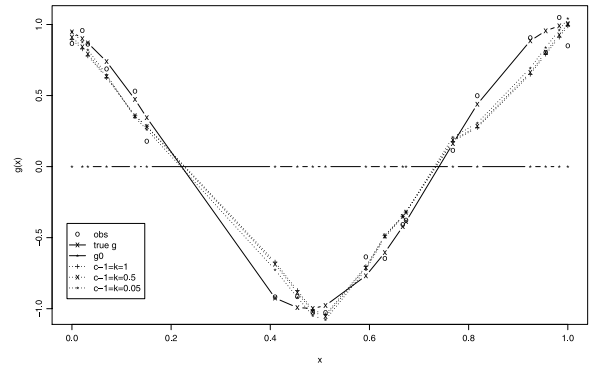

(a) Gaussian

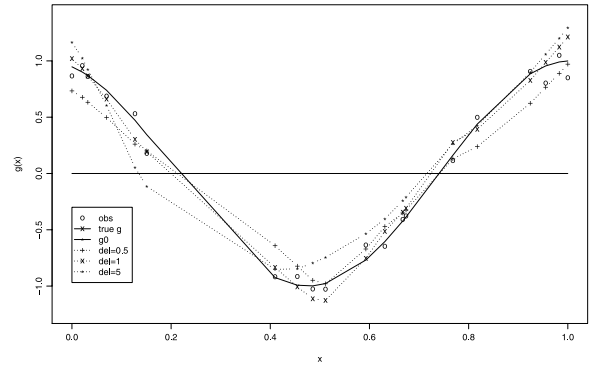

(b) Ellipsoid

FIG 3. *Wavelet smoother for $g(x) = \cos(2\pi x)$ and $g_0 \equiv 0$.*

### 6.1. Model checking

The model checking approach based on Bayes factors developed in the previous section has been tested on simulated examples. For this, we generated observations form Equation (1.1) with the true function $g(x_i) = \cos(2\pi x_i)$ where $x_i$, $i = 1, 2, \ldots, 20$ were sampled from $U(0, 1)$. For the error term, $\epsilon_i$, we used $\sigma^2 = 0.1$. Then, for illustration purposes, we considered three different $g_0$ functions in $(M_0 : g = g_0)$:

   (i) $g_0(x) = \cos(2\pi x)$,
  (ii) $g_0(x) = 4|x - 0.5| - 1$ and
 (iii) $g_0 \equiv 0$.

Note that, the $g_0$ function in (i) corresponds to the true function. The $g_0$ functions in both (i) and (ii) are similar while the last one is very different from the true function $g$. Therefore, it is fair to assume that the Bayes factor (see equation (5.1)) for the first two cases should not provide evidence against the model $M_0 : g = g_0$ while we can expect strong evidence in the case of the third function. These Bayes factors are given in Table 1. From this table, it can be seen that the model corresponding to the correct function ($g_0(x) = \cos(2\pi x)$) obtains the largest Bayes factor followed by that for $g_0(x) = 4|x - 0.5| - 1$. Moreover, if we test $M_0 : g_0(x) \equiv 0$ against $M_1 : g_0(x) \neq 0$, the Bayes factor favors $M_1$ with strong evidence.

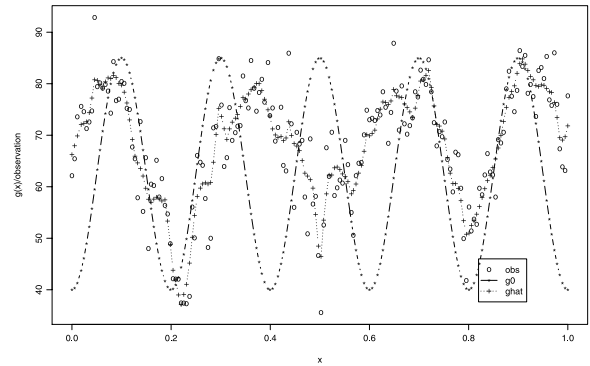

(a) Gaussian

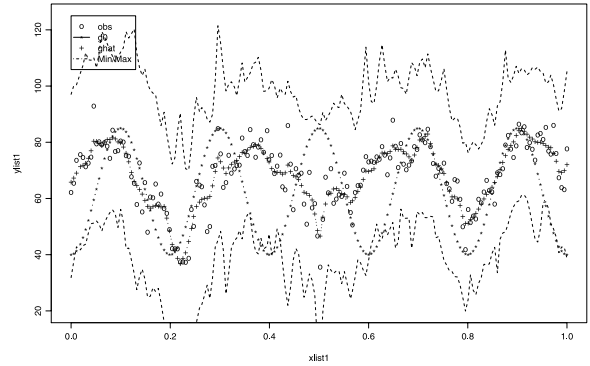

(b) Ellipsoid

Fig 4. *Wavelet smoother for the "Humidity data".*

Table 1
*Bayes factor for $M_0 : g = g_0$ vs $M_1 : g \neq g_0$*

| $g_0$ | $B_{01}(y)$ | Evidence |
|---|---|---|
| $\cos(2\pi x)$ | 933.4275 | very strongly favors $M_0$ |
| $4|x - 0.5| - 1$ | 57.4735 | strongly favors $M_0$ |
| 0 | $7.2845 \times 10^{-6}$ | very strongly favors $M_1$ |

## 7. Conclusions

In this paper we suggest a simple approach to nonparametric regression by proposing an alternative to dealing with complicated analyses on function spaces. The proposed technique uses fuzzy sets to quantify the available prior information on a function space by starting with a "prior guess" baseline regression function $g_0$. First, wavelet decomposition is used to represent both the unknown regression function $g$ as well as the prior guess $g_0$. Then the prior uncertainty of $g$ relative to its distance from $g_0$ is specified in the form of a membership function which translates this distance into a measure of distance between the corresponding wavelet coefficients. Furthermore, a Bayesian test is proposed to check whether the baseline function $g_0$ is compatible with the data or not.

**Acknowledgments.** We thank the editors and two referees for providing critical comments which have brought significant improvements to our presentation.